\documentclass[universe,article,accept,moreauthors,pdftex]{mdpi} 

\firstpage{1} 
\pubvolume{5}
\issuenum{8}
\articlenumber{186}
\pubyear{2019}
\copyrightyear{2019}
\history{Received: 8 May 2019; Revised: 7 August 2019; Accepted: 8 August 2019; Published: 12 August 2019}
\pdfoutput=1 

\usepackage[labelformat=simple]{subcaption}

\DeclareCaptionLabelFormat{subcaptionlabel}{\normalfont(\textbf{#2}\normalfont)}
\Title{Impact of the Nuclear Equation of State on the Stability of Hybrid Neutron Stars}

\Author{Mateusz Cierniak \textsuperscript{1}\textsuperscript{,}*\orcidA{}, Tobias Fischer \textsuperscript{1}\orcidB{}, Niels-Uwe Bastian \textsuperscript{1}\orcidC, Thomas Kl\"{a}hn \textsuperscript{2} \mbox{and Marc Salinas \textsuperscript{2}}}

\AuthorNames{Mateusz Cierniak, Tobias Fischer, Niels-Uwe Bastian, Thomas Klahn, }

\address{\textsuperscript{1} \quad Institute of Theoretical Physics, University of Wrocław

\textsuperscript{2} \quad Department of Physics and Astronomy, California State University Long Beach, 1250 Bellflower Blvd., \mbox{Long Beach, CA 90840, USA}}

\corres{Correspondence: mateusz.cierniak@uwr.edu.pl}

\abstract{We construct a set of equations of state (EoS) of dense and hot matter with a 1st order phase transition from a hadronic system to a deconfined quark matter state. In this two-phase approach, hadrons are described using the relativistic mean field theory with different parametrisations and the deconfined quark phase is modeled using vBag, a bag--type model extended to include vector interactions as well as a simultaneous onset of chiral symmetry restoration and deconfinement. This feature results in a non--trivial connection between the hadron and quark EoS, modifying the quark phase beyond its onset density. We find that this unique property has an impact on the predicted hybrid (quark core) neutron star mass--radius relations.
}



\begin{document}

\section{Introduction}

The current state-of-the-art in the description of strongly interacting matter is the theory of Quantum Chromodynamics (QCD). Its elementary degrees of freedom are quarks, spin 1/2 fermions with an electric charge of +2/3 or -1/3 and gluons, spin 1 bosons with an electric charge of 0 acting as strong interaction force carriers. Each of those particles has a chromodynamic charge called color. Given the QCD feature of running coupling, i.e., rapidly increasing interaction strength with respect to interaction range (cf. \cite{Roberts:2015lja}) and, due to the fact that only color-neutral particles are being observed in nature, it is believed that color charged particles cannot be separated and can only exist in bound states. This is called confinement. Additionally, QCD exhibits a phenomenon called dynamic chiral symmetry breaking ($\mathrm{D}\chi\mathrm{SB}$) along with chiral symmetry restoration at high densities and temperatures. This feature gives rise to approximately massless Nambu--Goldstone bosons (pions) as well as generates most of the quark mass. $\mathrm{D}\chi\mathrm{SB}$ is believed to account for roughly $\mathrm{98\%}$ of the mass of a~proton~\cite{Roberts:2015lja}.

The first-principle calculations of QCD (i.e., lattice QCD) produce reliable results for baryon chemical potentials up to about twice the given temperature (c.f.  \cite{Fodor:2004nz,Aoki:2006we,Bazavov:2017dus,Gunther:2016vcp,Bazavov:2018mes}). They predict a cross-over phase transition at the temperature of $\mathrm{156.5\pm1.5}$ MeV \cite{Bazavov:2018mes} to the deconfined and chiraly restored quark--gluon plasma (QGP). This transition was analyzed up to a baryon chemical potential of about 300 MeV \cite{Bazavov:2017dus,Gunther:2016vcp}. Astrophysical systems are characterized by high densities up to and beyond nuclear saturation and lower temperatures, reaching from 0 to not more than 60 MeV. Asymptotic freedom implies that a phase transition to the QGP will eventually occur. The nature of this transition, its~relevance for the study of such systems, and the QGP matter afterwards is one of the pressing issues of contemporary studies of compact stars.

Due to the unavailability of first-principle calculations of cold dense QCD, studies that require the description of dense strongly interacting matter have to rely on effective models. Such models should exhibit all the properties of QCD, i.e., chiral symmetry breaking (and restoration) as well as confinement. Currently, there are only a few approaches that attempt to consistently describe hadron matter at the level of quarks and gluons at high density (cf. for example \cite{Bastian:2018wfl, Dexheimer:2009hi, Steinheimer:2010ib, Marczenko:2017huu, Marczenko:2018jui}), with the majority of studies relying on models constructed using the two-phase approach \cite{Cleymans:1985wb}, i.e., with a separate hadron and quark matter models and a transition between them emulating the effect of deconfinement. A~general review of recent developments concerning the EoS in astrophysical applications can be found in \cite{Oertel:2017,Fischer:2017zcr}.

A common quark matter model for such studies is the tdBag model \cite{Farhi:1984qu}. It approximates the effect of quark confinement but does not exhibit $\mathrm{D}\chi\mathrm{SB}$ and lacks repulsive vector interactions necessary to reach the two solar mass neutron star limit in agreement with recent observations of PSRJ1614-2230 and PSRJ0348+0432 with masses of $\mathrm{1.928\pm 0.017}$ $\mathrm{M_{\odot}}$ \cite{Demorest:2010bx, Fonseca:2016tux} and $\mathrm{2.01\pm 0.04}$ $\mathrm{M_{\odot}}$ \cite{Antoniadis:2013pzd}, respectively. vBag, a~novel extension of the tdBag approach is an effective model for astrophysical studies with both vector repulsion, chiral symmetry restoration and deconfinement (cf. \cite{Klahn:2015mfa, Klahn:2016uce, Fischer:2016ojn, Cierniak:2018aet}). The two latter properties are assumed to be simultaneous. This is achieved via a correction to the quark EoS based on the hadron EoS at the chiral transition. The results of some Dyson--Schwinger studies justify this assumption in the cross-over domain of the QCD phase diagram; however, this is far from certain in the desired high density domain (cf. \cite{Qin:2010nq, Fischer:2014ata}), therefore it will be considered a model assumption of~vBag.

The manuscript is organized as follows. In Section \ref{S2}, we introduce vBag and its derivation from the Dyson--Schwinger equations (DSE) of QCD. In Section \ref{S5}, the temperature--density phase diagrams of vBag will be discussed and Section \ref{S4} will focus on the resulting hybrid neutron star mass--radius relations. A summary of the findings will be presented in Section \ref{Ssum}.

\section{vBag} \label{S2}

The general QCD in-medium single flavor quark propagator has the form~\cite{Rusnak:1995ex,Roberts:2000aa}

\begin{equation}
S^{-1}(p^{2},\tilde{p}_{4})=i\vec{\gamma}\vec{p}A(p^{2},\tilde{p}_{4})+i\gamma_{4}\tilde{p}_{4}C(p^{2},\tilde{p}_{4})+B(p^{2},\tilde{p}_{4})\/,
\end{equation}

\noindent with ${\tilde{p}_{4}=p_{4}+i}\mu$, 
and $\mu$ being the chemical potential. The gap functions A, B and C account for non-ideal behavior due to interactions. They can be derived by using the quark DSE,

\begin{equation}
S^{-1}(p^{2},\tilde{p}_{4})=i\vec{\gamma}\vec{p}+i\gamma_{4}\tilde{p}_{4}+m+\Sigma(p^{2},\tilde{p}_{4}),
\end{equation}

\noindent with the self-energy term

\begin{equation}
\Sigma(p^{2},\tilde{p}_{4})=\int\frac{d^{4}q}{(2\pi)^{4}}g^{2}(\mu)D_{\rho\sigma}(p-q,\mu)\frac{\lambda^{\alpha}}{2}\gamma^{\rho}S(q^{2},\tilde{q}_{4})\Gamma^{\sigma}_{\alpha}(q,p,\mu),
\end{equation}

\noindent where m is the bare mass, $D_{\rho\sigma}(p-q,\mu)$ is the dressed--gluon propagator and $\Gamma^{\sigma}_{\alpha}(q,p,\mu)$  is the dressed quark--gluon vertex. One solves this equation by imposing a specific set of approximations~\cite{Klahn:2015mfa} to the self energy term $\Sigma(p^{2},\tilde{p}_{4})$. The first of them is the so-called rainbow truncation \cite{GutierrezGuerrero:2010md}, the leading order in a systematic, symmetry-preserving DSE truncation scheme \cite{Munczek:1994zz,Bender:1996bb},  

\begin{equation}
\Gamma^{\sigma}_{\alpha}(q,p,\mu)=\frac{\lambda_{\alpha}}{2}\gamma^{\sigma}.
\end{equation}

The second is an effective gluon propagator which is set to be constant in momentum space up to a hard cut-off $\Lambda$,

\begin{equation}
g^{2}D_{\rho\sigma}(p-q,\mu)=\frac{1}{m^{2}_{G}}\Theta(\Lambda^{2}-\vec{p}^{2})\delta_{\rho\sigma}\/,
\end{equation}

\noindent equivalent to a quark-quark contact interaction in configuration space analogous to the Nambu-- Jona--Lasino model (NJL, cf. \cite{Buballa:2003qv}). The Heaviside function $\Theta$ provides a {three}-momentum cutoff for all momenta $\vec{p}^{2}>\Lambda^{2}$. $\Lambda$ represents a regularization mass scale which, in a realistic treatment, would be removed from the model by taking the limit $\Lambda\to\infty$. For the NJL model, this procedure fails and $\Lambda$ is typically used as a simple UV cutoff. Different regularization procedures are available; in~fact, the~regularization scheme does not
have to affect UV divergences only, e.g., infra-red (IR) cutoff schemes can be used to remove unphysical implications \cite{Ebert:1996vx}. 

The term $\mathrm{m_{G}}$ in the gluon propagator refers to the gluon mass scale and defines the coupling strength. These approximations allow us to derive the gap equations. The A gap function has a trivial $\mathrm{A=1}$ solution, the rest takes the form

\begin{equation}
B(p^{2},\tilde{p}_{4})=m+\frac{16N_{c}}{9m^{2}_{G}}\int_{\Lambda}\frac{d^{4}q}{(2\pi)^{4}}\frac{B(q^{2},\tilde{q}_{4})}{\vec{q}^{2}A^{2}(q^{2},\tilde{q}_{4})+\tilde{q}^{2}_{4}C^{2}(q^{2},\tilde{q}_{4})+B^{2}(q^{2},\tilde{q}_{4})},
\end{equation}

\begin{equation}
\tilde{p}^{2}_{4}C(p^{2},\tilde{p}_{4})=\tilde{p}^{2}_{4}+\frac{8N_{c}}{9m^{2}_{G}}\int_{\Lambda}\frac{d^{4}q}{(2\pi)^{4}}\frac{\tilde{p}_{4}\tilde{q}_{4}C(q^{2},\tilde{q}_{4})}{\vec{q}^{2}A^{2}(q^{2},\tilde{q}_{4})+\tilde{q}^{2}_{4}C^{2}(q^{2},\tilde{q}_{4})+B^{2}(q^{2},\tilde{q}_{4})},
\end{equation}
where $\int_{\Lambda}=\int\Theta(\vec{p}^{2}-\Lambda^{2})$. Both equations can be recast in terms of scalar and vector densities of an~ideal spin-degenerate Fermi gas,

\begin{equation}\label{eq8}
B=m+\frac{4N_{c}}{9m^{2}_{G}}n_{s}(\mu^{*},B),
\end{equation}

\begin{equation}\label{eq9}
\mu=\mu^{*}+\frac{2N_{c}}{9m^{2}_{G}}n_{v}(\mu^{*},B),
\end{equation}
where

\begin{equation}
n_{s}(\mu^{*},B)=2\sum_{\pm}\int_{\Lambda}\frac{d^{3}q}{(2\pi)^{3}}\frac{B}{E}\left(\frac{1}{2}-\frac{1}{1+exp(E^{\pm}/T)}\right),
\end{equation}

\begin{equation}
n_{v}(\mu^{*},B)=2\sum_{\pm}\int_{\Lambda}\frac{d^{3}q}{(2\pi)^{3}}\frac{\mp 1}{1+exp(E^{\pm}/T)},
\end{equation}
\textls[-33]{with ${E^{2}=\vec{p}^{2}+B^{2}}$ and ${E^{\pm}=E~\pm~}\mu^{*}$.
The integrals have no explicit external momentum dependence ($p$), therefore the gap solutions are constant for a given $\mu$.  Typically, for DSE calculations, the pressure is determined in the steepest descent approximation. It consists of an ideal Fermi gas and interaction~contributions:}

\begin{equation}
P_{FG}=TrLnS^{-1}=2N_{c}\int_{\Lambda}\frac{d^{4}q}{(2\pi)^{4}}Ln\left(\vec{p}^{2}+\tilde{p}^{2}_{4}+B^{2}\right),
\end{equation}

\begin{equation}
P_{I}=-\frac{1}{2}Tr\Sigma S=\frac{3}{4}m^{2}_{G}\left(\mu-\mu^{*}\right)^{2}-\frac{3}{8}m^{2}_{G}\left(B-m\right)^{2}.
\end{equation}

The merit of the NJL model is the ability to describe chiral symmetry breaking as the formation of a scalar condensate and the restoration of chiral symmetry as melting of the same. The chosen hard cutoff scheme reproduces standard NJL model results and allows for describing quarks as a quasi ideal gas of fermions (assuming constant mass equal to the quarks bare mass) shifted by a constant factor (denoted as $B_{\chi,f}$), as seen in Figure 1 of \cite{Klahn:2015mfa}. This is similar to the standard tdBag model approach (cf.~\cite{Farhi:1984qu}). Therefore, we can express the single-flavor pressure as

\begin{equation}\label{eq14}
P_{f}(\mu_{f})=P_{FG,f}(\mu^{*}_{f})+\frac{K_{v}}{2}n^{2}_{FG,f}(\mu^{*}_{f})-B_{\chi,f}.
\end{equation}

\textls[-15]{The second term corresponds to the vector condensate, where ${K_{v}}$ relates to the vector current--current interaction coupling constant. In our approach, it is defined in terms of the gluon mass scale:}

\begin{equation}
K_{v}=\frac{2}{9m^{2}_{G}}.
\end{equation}

In combination with the modification of the effective chemical potential $\mu^{*}$, it causes stiffening of the EoS with increasing density. This effect is illustrated in {Figure 2} of \cite{Cierniak:2018aet}. 
  
From Equation (\ref{eq8}), it is evident that a corresponding scalar current--current interaction coupling constant can be defined as ${K_{s}=2K_{v}}$. The relation of the coupling constants is consistent with the result obtained after Fierz transformation of the one-gluon exchange interaction \cite{Buballa:2003qv}. However, we~absorbed the effect of scalar interactions in $B_{\chi}$ and vary ${K_{v}}$ as an independent model parameter. This~procedure is common for NJL-type model studies. Taking the vector interaction into account then results in a~modification of the effective chemical potential $\mu^{*}$ and pressure as evident \mbox{from Equations (\ref{eq9}) and (\ref{eq14}).} This term is not included in the standard tdBag model. Chiral symmetry is restored when $P_{f}(\mu_{f})>0$, and therefore the critical chemical potential can be defined as 

\begin{equation}
P_{f}(\mu_{\chi,f})=0.
\end{equation} 

For two-flavor quark matter, this condition is redefined as
\begin{equation} \label{eq17}
\tilde{P}^{Q}(\mu_{\chi})=\sum_{f}P_{f}(\mu_{f})=0
\end{equation} 
to avoid sequential chiral symmetry restoration. This is done so that we can impose simultaneous chiral symmetry restoration and deconfinement, which can be achieved by exploiting the fact that the total pressure is fixed only up to a constant. Therefore, we can impose 

\begin{equation}
P^{Q}=\sum_{f}P_{f}+B_{dc}.
\end{equation}

By defining ${B_{dc}}$ as the hadron pressure at the point of quark chiral symmetry restoration, one~easily sees that both transitions would coincide. We can now write the full set of equations that define~vBag

\begin{equation}
\mu_{f}=\mu^{*}_{f}+K_{v}n_{FG,f}(\mu^{*}_{f}),
\end{equation}

\begin{equation}
n_{f}(\mu_{f})=n_{FG,f}(\mu^{*}),
\end{equation}

\begin{equation}
P_{f}(\mu_{f})=P_{FG,f}(\mu^{*}_{f})+\frac{K_{v}}{2}n^{2}_{FG,f}(\mu^{*}_{f})-B_{\chi,f},
\end{equation}

\begin{equation}
\epsilon_{f}(\mu_{f})=\epsilon_{FG,f}(\mu^{*}_{f})+\frac{K_{v}}{2}n^{2}_{FG,f}(\mu^{*}_{f})+B_{\chi,f},
\end{equation}

\begin{equation}
P^{Q}=\sum_{f}P_{f}+B_{dc},
\end{equation}

\begin{equation}
\epsilon^{Q}=\sum_{f}\epsilon_{f}-B_{dc},
\end{equation}
where $\epsilon$ denotes energy density and n is the particle number density. 

Since the vBag EoS is defined in terms of a free gas of quasiparticle fermions, its temperature extension follows by simply taking the full temperature-dependent terms of this free Fermi gas instead of its ground state (cf. \cite{Klahn:2016uce}), with
\begin{equation}
\frac{\partial P}{\partial T}=s
\end{equation}
being the entropy density of the system. In considering isospin asymmetry, it is useful to introduce the baryon and charge chemical potentials

\begin{equation}
\mu_{B}=\mu_{u}+2\mu_{d},
\end{equation}
\begin{equation}
\mu_{C}=\mu_{u}-\mu_{d},
\end{equation}
and the associated densities
\begin{equation}
\frac{\partial P}{\partial\mu_{B}}=n_{B},
\end{equation}
\begin{equation}
\frac{\partial P}{\partial\mu_{C}}=n_{C}.
\end{equation}

Furthermore, the assumption of isospin-independent chiral symmetry restoration (Equation (\ref{eq17})) allows us to define the (critical) chiral baryon chemical potential $\mu_{B\chi}$, such that
\begin{equation} \label{eq30}
\tilde{P}^{Q}(T,\mu_{B\chi},\mu_{C})=\sum_{f}P_{f}(T,\mu_{f})=0.
\end{equation}

The deconfinement bag constant ${B_{dc}}$ is therefore defined in term of $\mu_{B\chi}$ as
\begin{equation} \label{eq31}
B_{dc}=P^{H}(T,\mu_{B\chi},\mu_{C}).
\end{equation}

The treatment of chiral symmetry restoration and deconfinement as simultaneous allows us to keep the description of chiral physics on the quark side of the EoS while avoiding the problem of using a nuclear EoS insensitive to the quark chiral restoration. This, however, results in ${B_{dc}}$ being a~function of both temperature and charge chemical potential. Thermodynamic consistency requires that additional terms be included in the expressions for charge density and entropy to account for this fact (as well as $\epsilon$, due to the Euler equation, cf. \cite{Klahn:2016uce}), i.e.,
\begin{equation}
\frac{\partial P}{\partial T}=s(T,\mu_{B},\mu_{C})=\sum_{f}s_{FG,f}(T,\mu_{B},\mu_{C})+\frac{\partial B_{dc}}{\partial T}:=\tilde{s}+s_{dc},
\end{equation}
\begin{equation}
\frac{\partial P}{\partial \mu_{C}}=n_{C}(T,\mu_{B},\mu_{C})=\sum_{f}Q_{f}n_{f}(T,\mu_{B},\mu_{C})+\frac{\partial B_{dc}}{\partial \mu_{C}}:=\tilde{n}_{C}+n_{C,dc},
\end{equation}
\begin{equation}
\epsilon^{Q}=\sum_{f}\epsilon_{f}-B_{dc}+Ts_{dc}+\mu_{Q}n_{Q,dc},
\end{equation}
where ${Q_{f}}$ is the charge of the given quark flavor (${Q_{u}=2/3}$, ${Q_{d}=-1/3}$) while ${s_{dc}}$ and ${n_{C,dc}}$ are functions of both quark and hadron equations of state.
\begin{equation} \label{eq35}
s_{dc}=s^{H}(T,\mu_{B\chi},\mu_{C})-\tilde{s}^{Q}(T,\mu_{B\chi},\mu_{C})\frac{n^{H}_{B}(T,\mu_{B\chi},\mu_{C})}{n^{Q}_{B}(T,\mu_{B\chi},\mu_{C})},
\end{equation}
\begin{equation} \label{eq36}
n_{C,dc}=n^{H}_{C}(T,\mu_{B\chi},\mu_{C})-\tilde{n}^{Q}_{C}(T,\mu_{B\chi},\mu_{C})\frac{n^{H}_{B}(T,\mu_{B\chi},\mu_{C})}{n^{Q}_{B}(T,\mu_{B\chi},\mu_{C})}.
\end{equation}

This non--trivial relation is not present in any standard two-phase equations of state and is unique to vBag. Obviously, the impact on the deconfined quark matter is largely dependent on the choice of hadronic EoS and will be explored in the following chapter.

\section{The Phase Diagram} \label{S5}

Having an equation of state dependent on both the baryon and charge chemical potential, as~well as temperature, one can compare the resulting phase diagrams of hot and dense matter. For this purpose, two hadronic equations of state based on the relativistic mean field theory (denoted DD2 \cite{Typel:2009sy} and NL3 \cite{Lalazissis:1996rd}) have been selected. The choice of these EoS was made such that they produce sufficiently different thermodynamic quantities for given chemical potentials. A summary of their properties can be found in Table \ref{tab1}.

\begin{table}[H]
\caption{Selected properties of the hadronic equations of state ($\mathrm{T=0}$, $\mu_{C}=0$) and their experimental constraints (cf. \cite{Motohiro:2015taa, Lattimer:2012xj, Kortelainen:2011ft}).}
\centering
\tablesize{\footnotesize}
\begin{tabular}{cccccc}
\toprule
\textbf{EoS}	& \textbf{Saturation Density} & \textbf{Binding Energy per Nucleon} & \textbf{Incompressibility} & \textbf{Symmetry Energy} & \textbf{Slope Par.}\\
& \boldmath{$\mathrm{n_{0}}$ $\mathrm{[fm^{-3}]}$} & \boldmath{$\mathrm{(E/A)_{\infty}}$ $\mathrm{[MeV]}$} & \boldmath{$\mathrm{K_{\infty}}$ $\mathrm{[MeV]}$} & \boldmath{$\mathrm{S_{v}}$ $\mathrm{[MeV]}$} & \boldmath{$\mathrm{L}$ $\mathrm{[MeV]}$}\\
\midrule
DD2 & $\mathrm{0.149065}$ & $\mathrm{16.02}$ & $\mathrm{242.7}$ & $\mathrm{32.73}$ & $\mathrm{57.94}$\\
NL3 & $\mathrm{0.1483}$ & $\mathrm{16.299}$ & $\mathrm{271.8}$ & $\mathrm{37.4}$ & $\mathrm{118.49}$\\
\hline
ex. & 0.158--0.159 & 15.8--16.2 & 220--260 & 29--32.7 & 40.5--61.9\\
\bottomrule
\end{tabular}
\label{tab1}
\end{table}

As evident from Figure \ref{fig3}, the quark matter EoS differ depending on the choice of hadronic EoS. This difference is constant for a given temperature and charge chemical potential, as to be expected from Equations \eqref{eq31},  \eqref{eq35} and  \eqref{eq36}. It is also evident from Figure \ref{fig4} that the hadronic EoS and resulting correction terms have no impact on the critical baryon chemical potential. This is imposed by Equation~ \eqref{eq30}.

The biggest difference in the phase diagram can be seen in Figure \ref{fig5}. The difference in the mixed phase onset density is to be expected given the change in stiffness of the hadronic EoS. The overall shape of the mixed phase is a direct result of the temperature dependence of ${B_{dc}}$ (i.e., ${s_{dc}}$).

\begin{figure}[H]
\centering
\begin{subfigure}{0.6\textwidth}
\includegraphics[width=\linewidth]{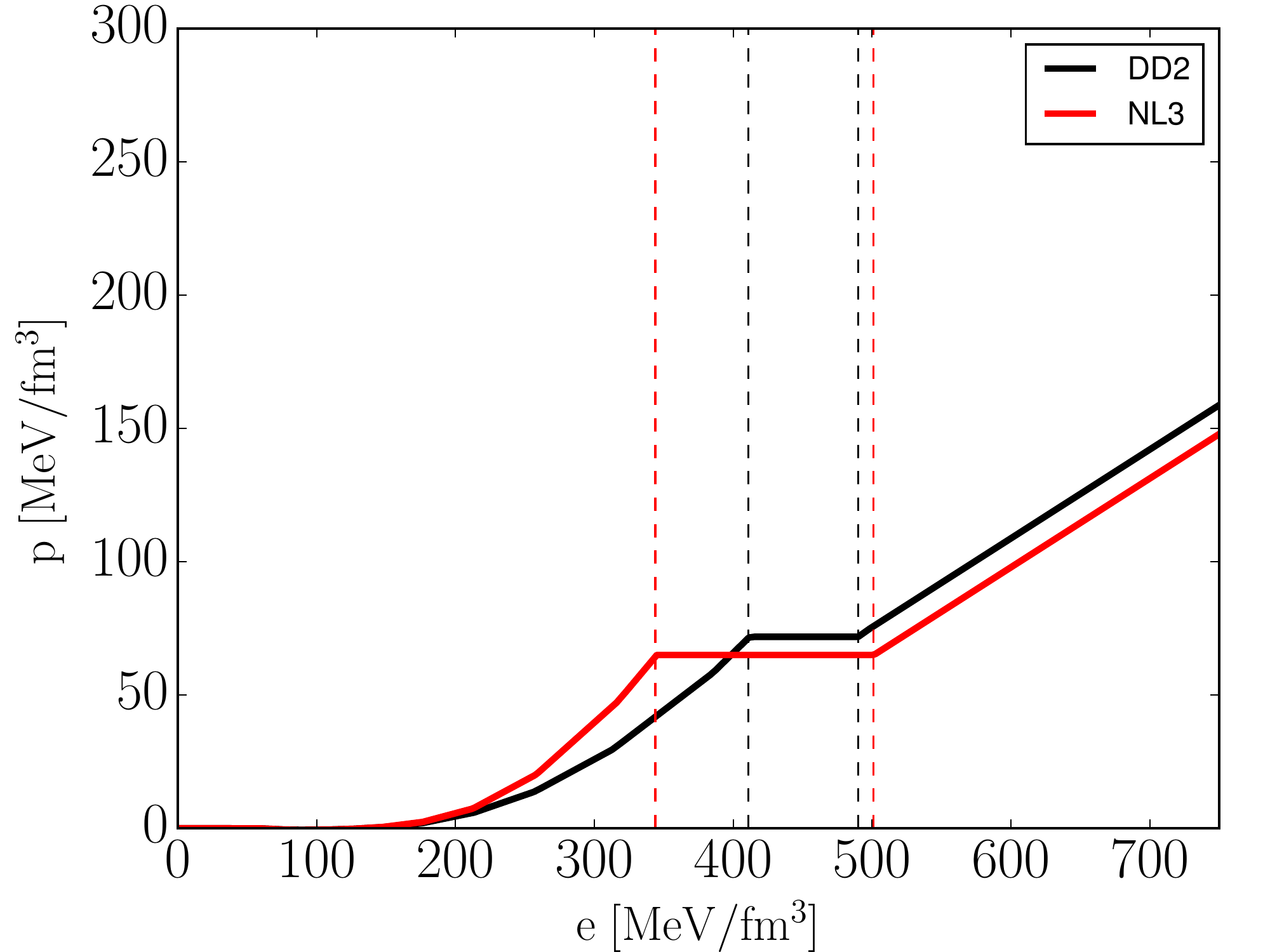}
\caption{}
\end{subfigure}
\caption{\textit{Cont}.}
\end{figure}

\begin{figure}[H]\ContinuedFloat
\centering
\setcounter{subfigure}{1}
\begin{subfigure}{0.6\textwidth}
\vspace{0.3cm}
\includegraphics[width=\linewidth]{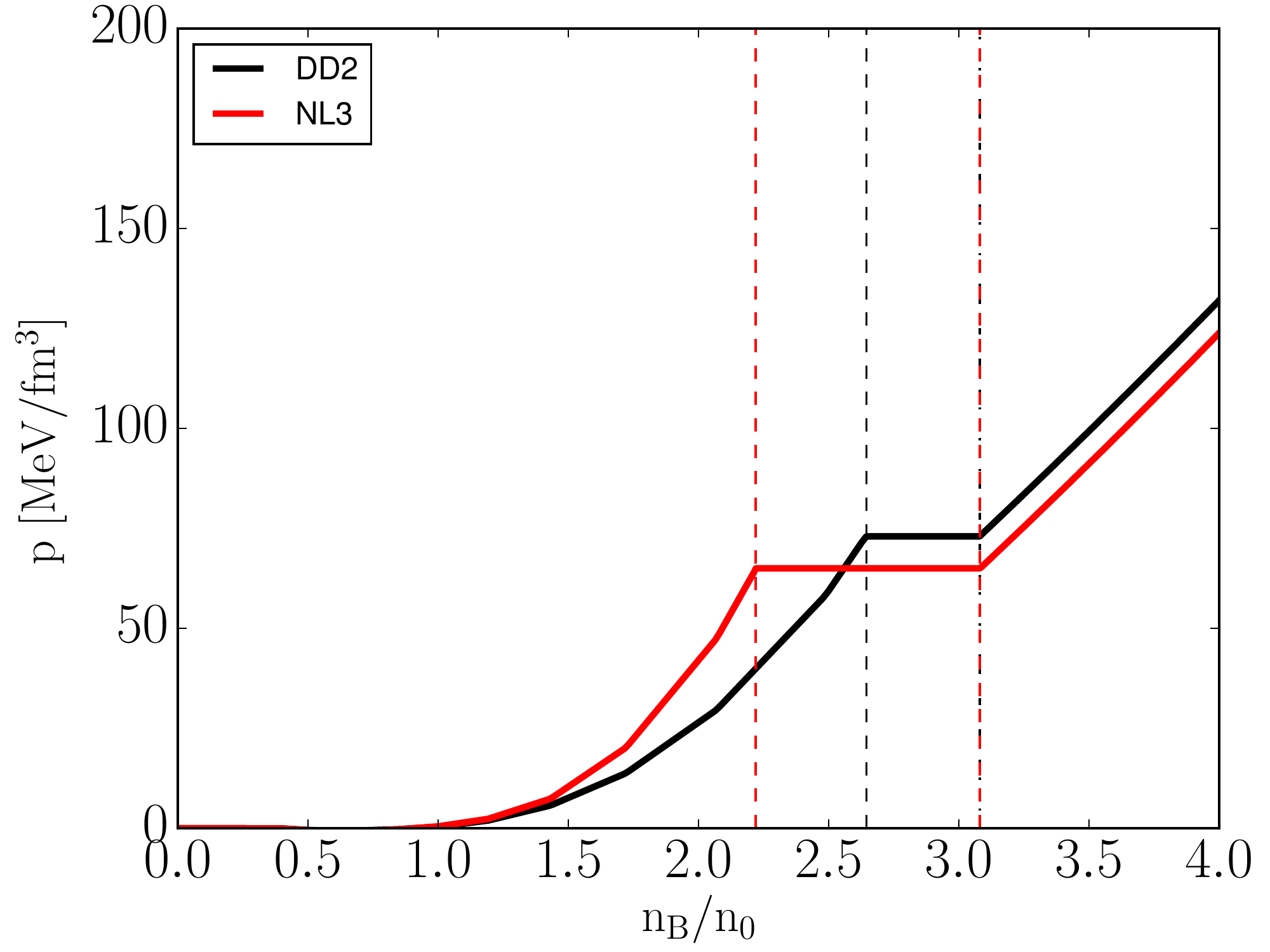}
\caption{}
\end{subfigure}
\caption{Pressure as a function of energy density (\textbf{a}) and baryon density (\textbf{b}) for cold isospin symmetric matter ($\mu_{C}=0$) modeled using vBag ($B^{1/4}_{\chi}=152.7$ MeV, $K_{v}=6\times 10^{-6}$ $\mathrm{MeV^{-2}}$) with different hadronic EoS. Dashed lines highlight mixed phase regions.}
\label{fig3}
\end{figure}
\unskip

\begin{figure}[H]
\centering
\includegraphics[width=0.6\linewidth]{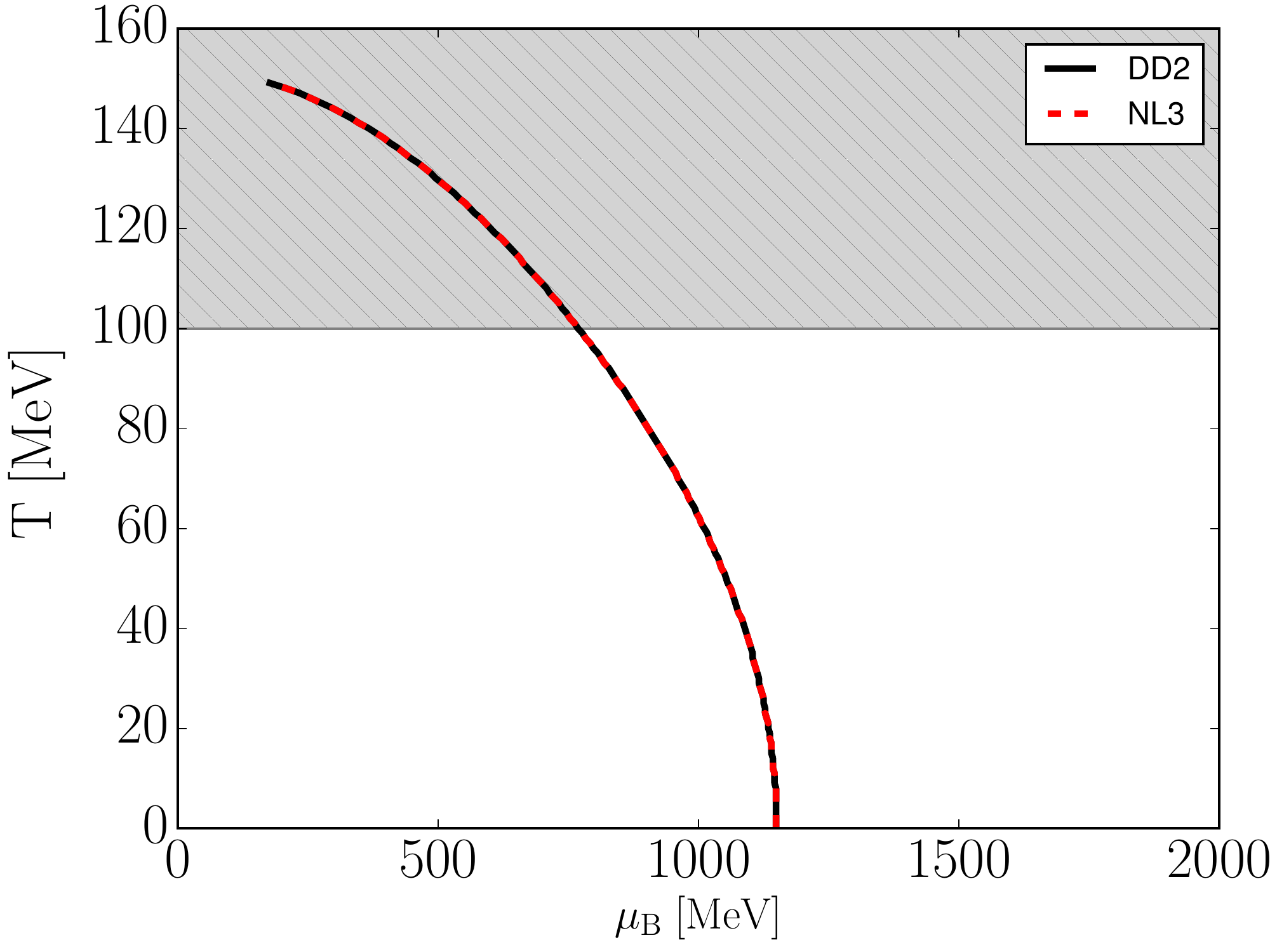}
\caption{The $T$--$\mu_{B}$ phase space of vBag ($B^{1/4}_{\chi}=152.7$ MeV, $K_{v}=6\times 10^{-6}$ $\mathrm{MeV^{-2}}$) for $\mu_{C}=0$. The~grey region lies outside of the models expected applicability domain. See text for details.}
\label{fig4}
\end{figure}

One needs to recognize that these phase diagrams cannot be considered accurate at temperatures of above 100 MeV \cite{Klahn:2016uce}. This is caused by $B_{\chi}$ (the quark scalar condensate) and its temperature dependence, which at an excess of a 100 MeV is no longer negligible. On top of that, the phase transition is 1st order by construction, while at vanishing densities this transition should become a~smooth crossover. These limitations are of no concern for static neutron star mass--radius relations; however, this might not be the case for neutron star mergers and therefore one needs to keep them in mind in future studies of neutron star phenomenology.

\begin{figure}[H]
\centering
\includegraphics[width=0.6\linewidth]{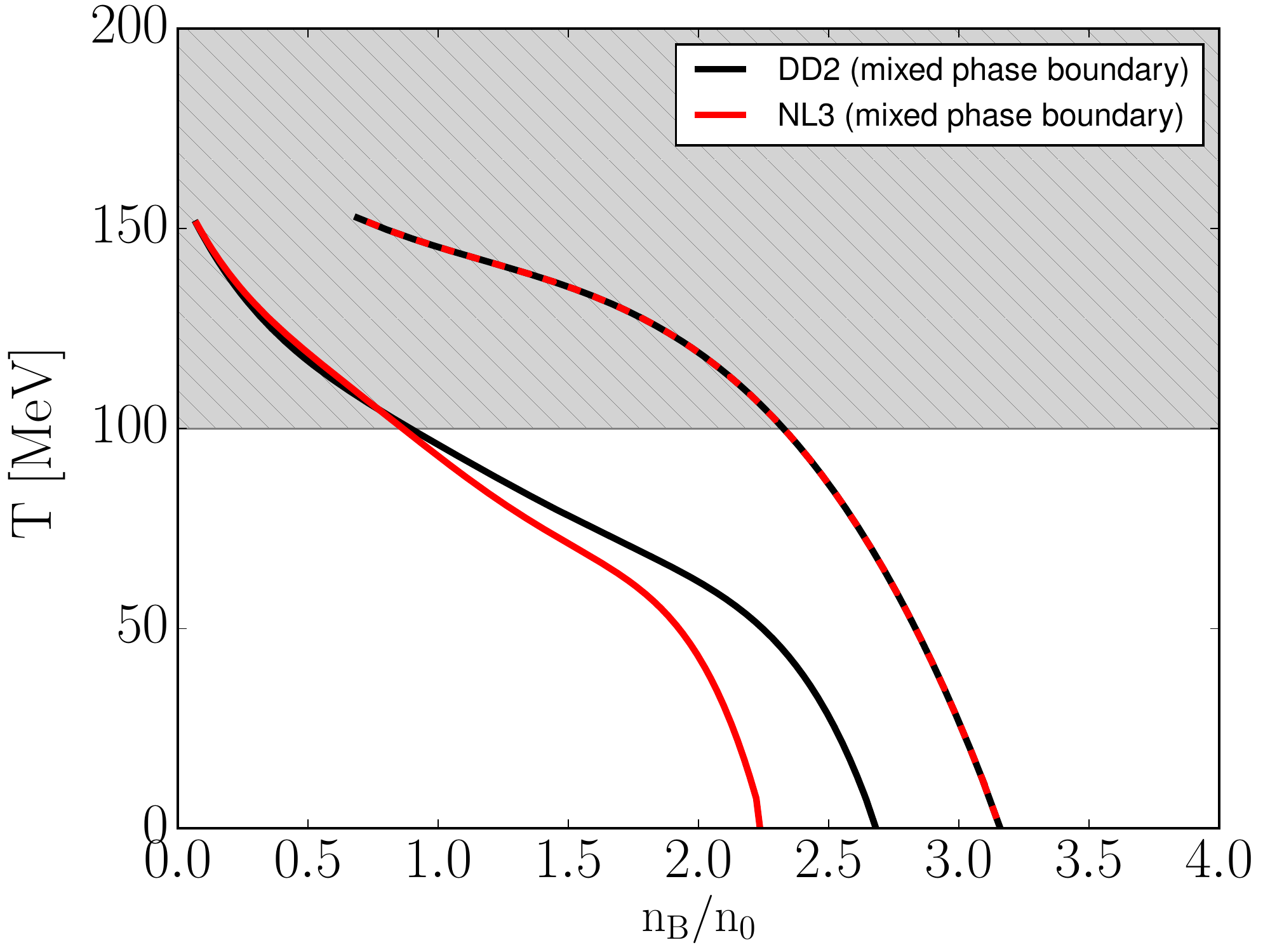}
\caption{The $T$--$n_{B}$ phase space of vBag ($B^{1/4}_{\chi}=152.7$ MeV, $K_{v}=6\times 10^{-6}$ $\mathrm{MeV^{-2}}$) for $\mu_{C}=0$. The~grey region lies outside of the models expected applicability domain. See text for details.}
\label{fig5}
\end{figure}

\begin{figure}[H]
\centering
\includegraphics[width=0.49\linewidth]{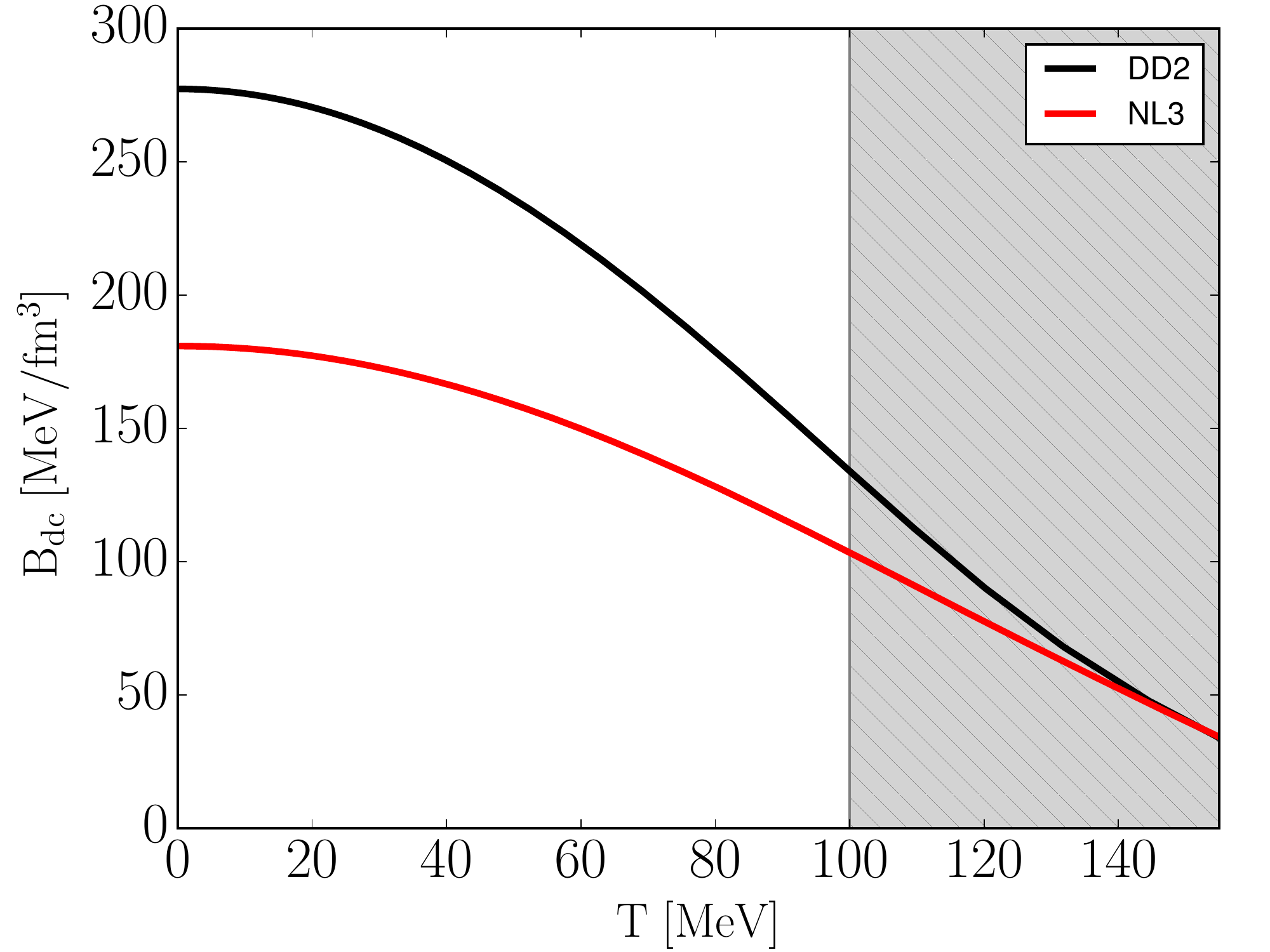}
\includegraphics[width=0.49\linewidth]{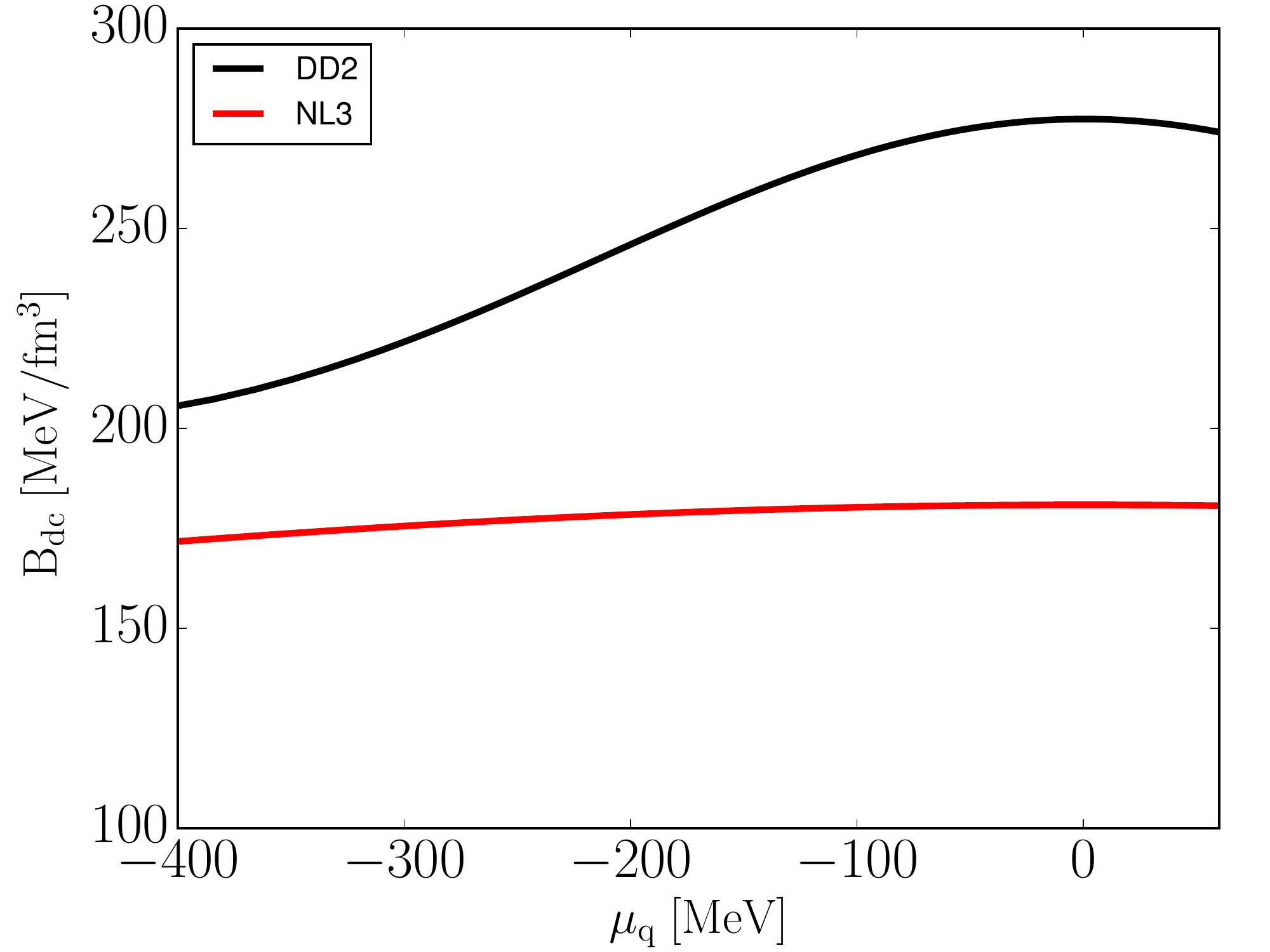}
\caption{The deconfinement bag constant $B_{dc}$ (Equation \eqref{eq31}) as a function of temperature for $\mu_{c}=0$ (\textbf{left}) and the charge chemical potential for $T=0$ (\textbf{right}). The~grey region lies outside of the models expected applicability domain. See text for details.}
\label{newfig}
\end{figure}

In the region below $\mathrm{T=100}$ MeV, we can clearly see significant differences in the phase structure of our model resulting from the choice of hadronic EoS (as evident from Figure \ref{newfig}). This fact is expected to have an impact on the derived astrophysical observables.

\section{Hybrid Neutron Stars} \label{S4}

\begin{figure}[H]
\centering
\begin{subfigure}{0.5\textwidth}
\includegraphics[width=\linewidth]{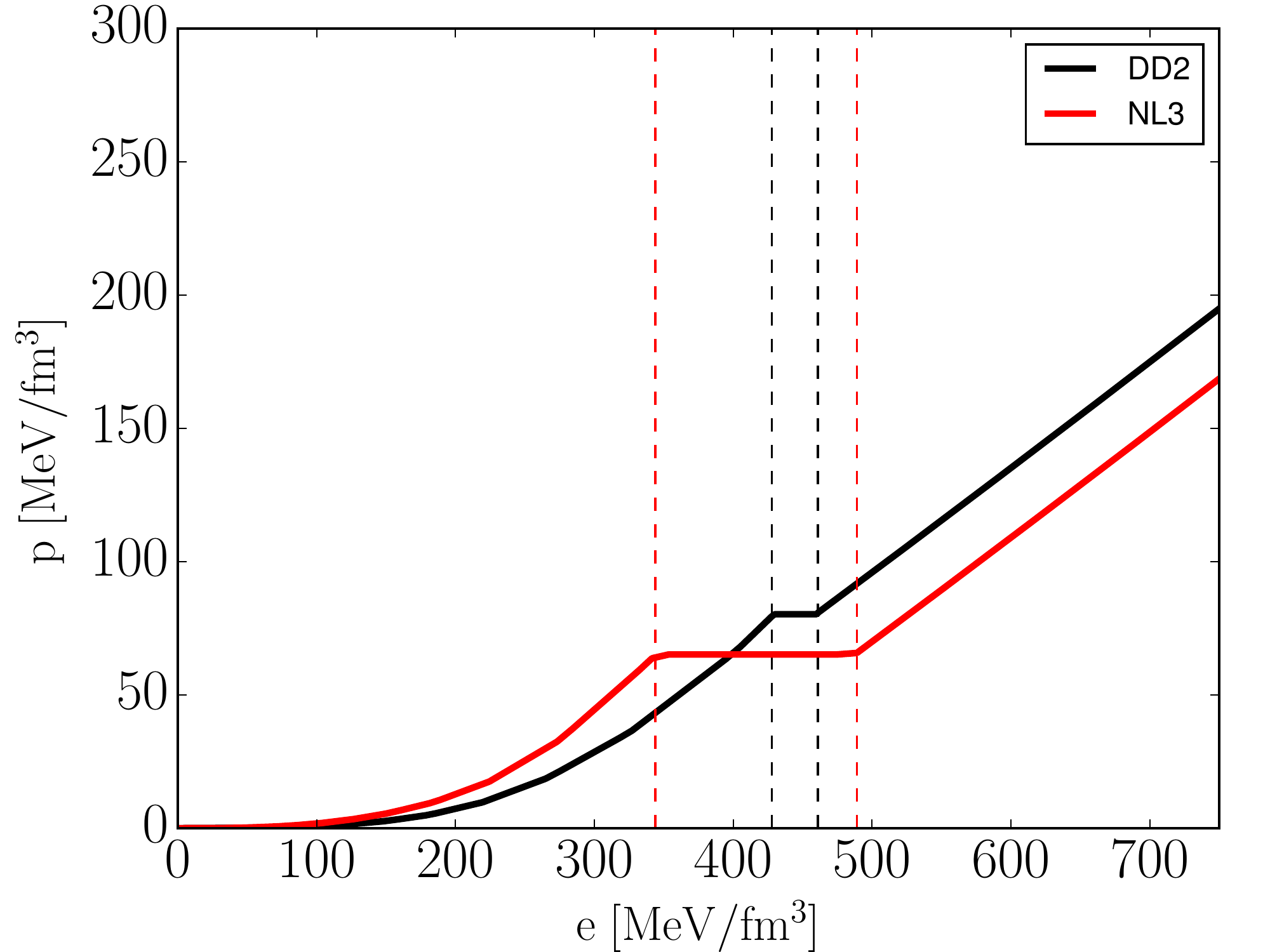}
\caption{}
\end{subfigure}
\begin{subfigure}{0.5\textwidth}
\vspace{0.3cm}
\includegraphics[width=\linewidth]{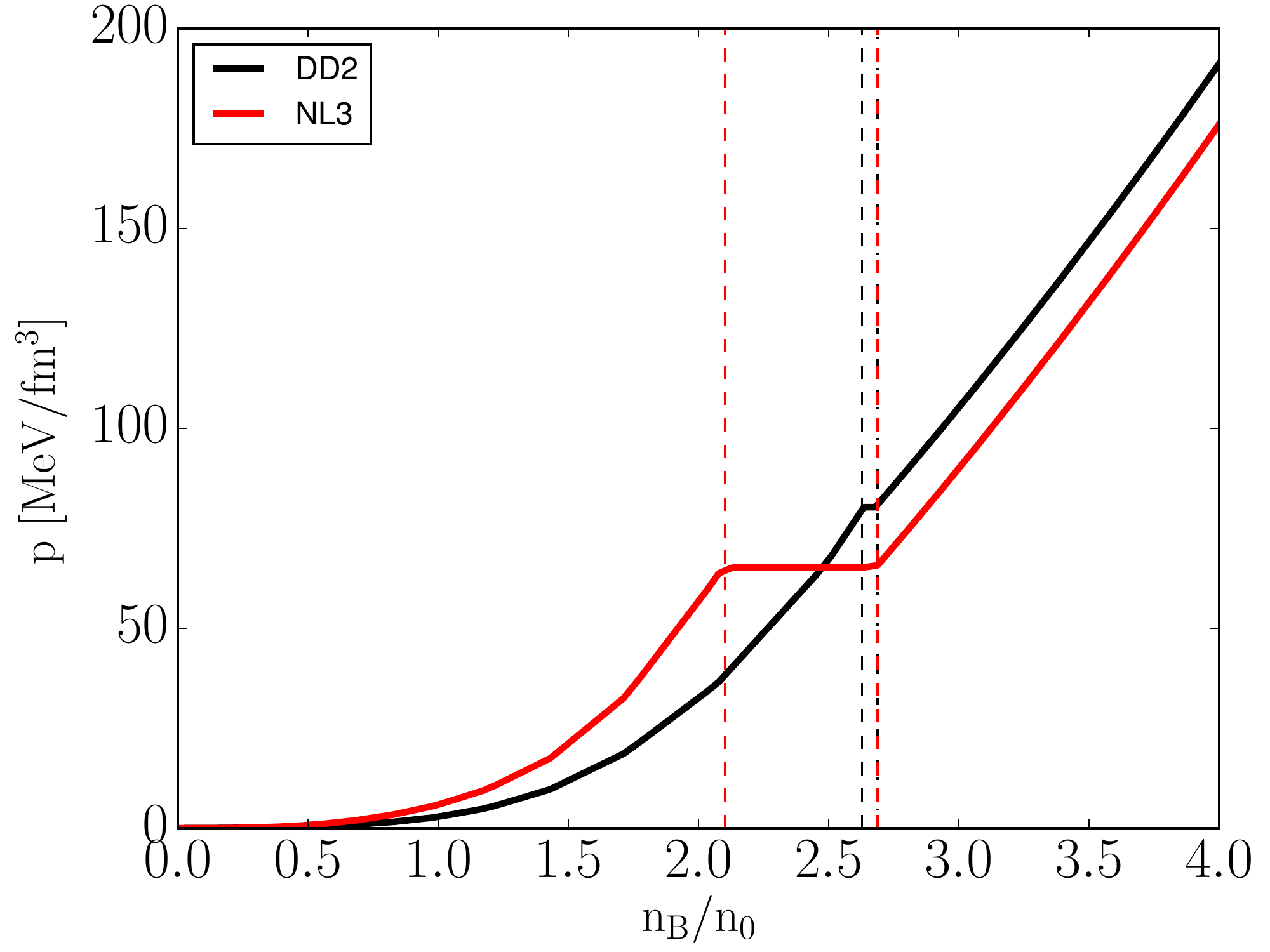}
\caption{}
\end{subfigure}
\caption{Pressure as a function of energy density (a) and baryon density (b) for cold dense matter in $\beta$-equilibrium modeled using vBag ($B^{1/4}_{\chi}=152.7$ MeV, $K_{v}=6\times 10^{-6}$ $\mathrm{MeV^{-2}}$) with different hadronic EoS. Dashed lines highlight mixed phase regions.}
\label{fig6}
\end{figure}

In order to derive the mass-radius relations of neutron stars, we need to consider dense matter under $\beta$-equilibrium and charge neutrality. These conditions will result in an EoS with strong isospin asymmetry (shown in Figure \ref{fig6}).

\begin{figure}[H]
\centering
\includegraphics[width=0.6\linewidth]{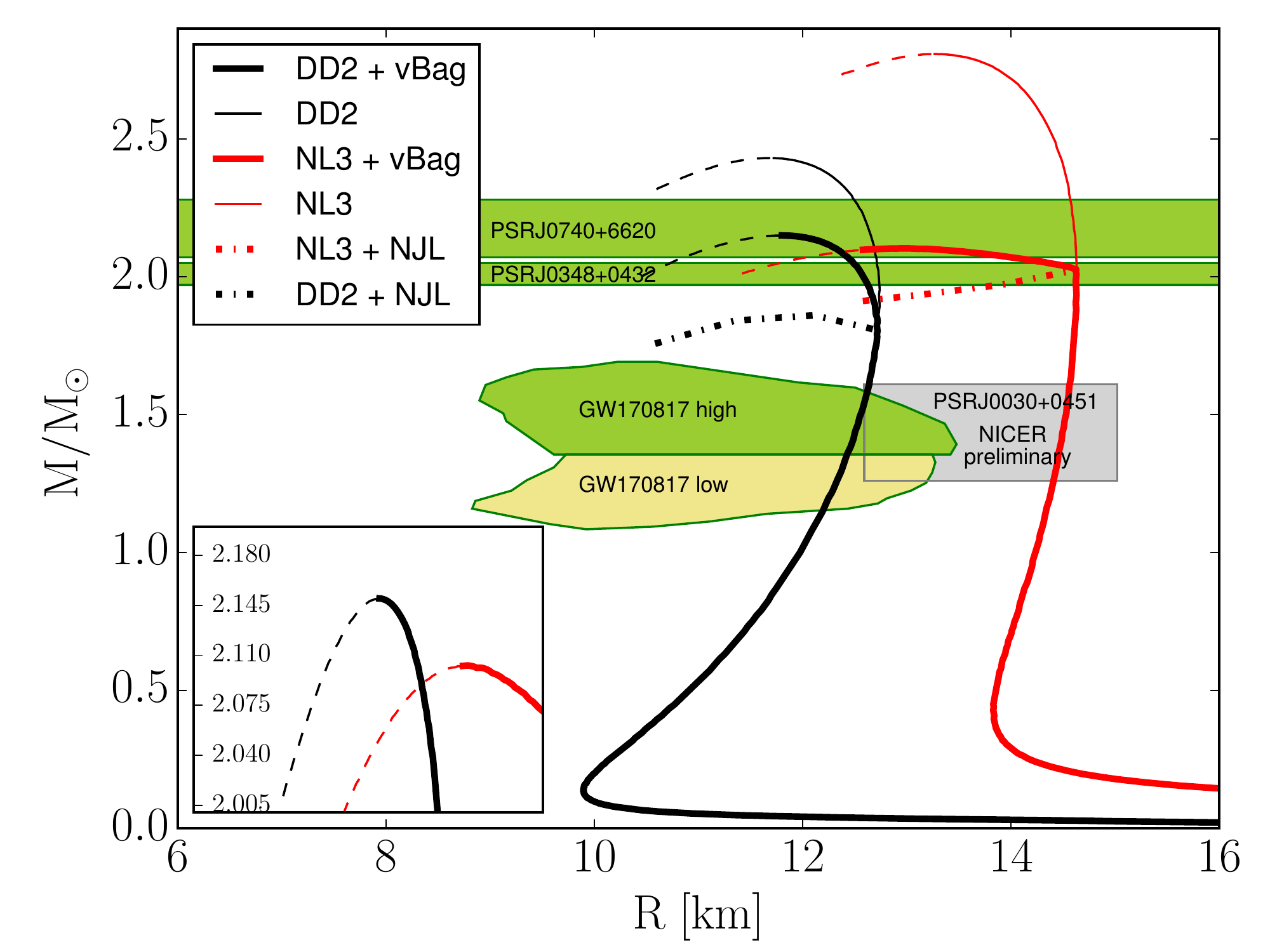}
\caption{\textls[-25]{Neutron star mass--radius curves obtained using vBag ($B^{1/4}_{\chi}=152.7$ MeV, $K_{v}=6\times 10^{-6}$ $\mathrm{MeV^{-2}}$) with two different hadronic EoS. The lower green, yellow and grey regions represent the mass--radius constraints of the neutron star merger event GW170817 for high and low mass posteriors (cf. \cite{Abbott:2018exr,De:2018uhw}) and the preliminary radius measurement of PSRJ0030+0451 from the NICER experiment \cite{NICER}, respectively. The~topmost green bands represent the mass measurements of the neutron star PSRJ0348+0432 \cite{Arzoumanian:2017puf} and the recent PSRJ0740+6620 \cite{Cromartie:2019kug}. Thin lines represent neutron stars with a purely hadronic core. Thick lines are hybrid neutron stars. Dashed lines show unstable mass--radius configurations. Dash-dotted lines represent hybrid EoS for $B_{dc}=0$ (equivalent to a pure NJL model, cf.\cite{Klahn:2015mfa}). The vector interaction strength $K_v$ was kept constant between all hybrid branches.}}
\label{fig7}
\end{figure}

The resulting mass--radius curves shown in Figure \ref{fig7} illustrate the impact of the underlying hadronic EoS on the quark matter in neutron stars. While Figure \ref{fig5} indicates one should expect different onset densities for hybrid stars (and also different onset star masses, due to the stiffness of the hadronic EoS), what is unexpected is the difference in maximum neutron star masses. Given that the free parameters of vBag are kept constant (specifically the vector repulsion strength $\mathrm{K_{v}}$, which directly impacts the maximum mass of stable neutron stars), the quark phase of DD2+vBag would not be able to support a heavier neutron star without the modification caused by ${B_{dc}}$ and ${n_{C,dc}}$.

\section{Conclusions} \label{Ssum}

The interior of neutron stars attains conditions that are presently inaccessible in nuclear physics and heavy-ion collision experiments. These astrophysical objects, together with core-collapse supernovae~\cite{Fischer:2017zcr} and binary neutron star mergers~\cite{Bauswein:2019}, enable us to probe the possible existence of {\em exotic} phases of dense matter such as the transition from ordinary nuclear (in general hadronic) matter to the quark--gluon plasma. How this quark--gluon plasma behaves in the high density domain is largely uncertain. The current state-of-the-art in dense matter modeling is a two-phase approach, which, at its core, is inconsistent due to the separate treatment of hadronic and quark degrees of freedom. Nevertheless, since the unified approach to the hadron--quark matter EoS is still not available based on QCD degrees of freedom at high baryon density, in order to study the potential impact of a~first-order hadron--quark phase transition at the interior of neutron stars, the two-phase approach has been commonly employed.

Furthermore, such a two-phase approach enables us to quantify the EoS in light of recent observations (see Figure \ref{fig7}). In particular, we focus on the recently available constraints obtained from the analysis of GW170817, which seems to favour not too large neutron star radii, and the preliminary data of NICER, which seem to agree with the latter only marginally. Nevertheless, both~of them are likely to have only a minor impact on properties of hybrid stars with hadron--quark transition densities corresponding to neutron star masses in excess of 1.5~M$_\odot$. In fact, within vBAG, the latter is a~direct consequence in order to obtain hybrid stars with a maximum mass of $\sim$2.1~M$_\odot$ or above. Note~that, within our setting, twin solutions were not obtained~\cite{Benic:2015}. Here, we select two vastly different hadronic EoS, DD2 and NL3, both of which belong to the class of relativistic mean-field models, which~result in very different radii of low- and intermediate-mass neutron stars (entirely determined by the hadronic EoS) but rather similar maximum masses though not identical. The differences arise exclusively from different correction terms related with $B_{\rm dc}$. It remains to be explored by future observations, in~particular the confirmed analysis of NICER data also for more massive pulsars on the order of 2~M$_\odot$, if~it will become eventually possible to further constrain the supersaturation density EoS in the light of the appearance of exotic phase at high densities.

Another aspect of dense matter under the conditions found in the interior of neutron stars is the onset of strangeness, expected to occur already at supersaturation density in hadronic \mbox{matter~(cf. \cite{Schaffner-Bielich:1996,Weissenborn:2012,Oertel:2017,Fortin:2018}~and} references therein).~Since hyperons are more massive than baryons, the~appearance of hyperons tends to soften the EoS with increasing density, which gave rise to the so-called hyperon puzzle in order to explain the existence of massive neutron stars with 2~M$_\odot$. However, little is presently known about hyperon--nucleon as well as hyperon--hyperon interactions. In particular, the~repulsive vector interaction may 'cure' the hyperon puzzle. Another solution has been proposed in terms of hadron--quark phase transition before hyperon degrees of freedom could substantially soften the EoS. However, studying this interesting scenario self-consistently requires not only strangeness as an explicit degree of freedom in the deconfined quark matter phase as was done in \cite{Klahn:2015mfa} but also in the hadronic EoS. This will allow us also to explore the hypothesis of absolutely stable strange matter~\cite{Witten:1984rs,Haensel:1986qb}, which has to be reviewed taking the restoration of chiral symmetry into account~\cite{Klahn:2015mfa}. However, this extends beyond the scope of the present paper. We leave the exploration of strangeness for a future more focused study.

\vspace{6pt} 

\authorcontributions{Conceptualization, M.C. and T.F; data curation, M.C and N.-U.B.; formal analysis, M.C.; investigation, M.C.; methodology, M.C.; project administration, M.C.; resources, M.C., T.F., N.-U.B., T.K. and M.S.; software, M.C., N.-U.B., T.K. and M.S.; supervision, T.F.; validation, M.C., T.F., N.-U.B., T.K. and M.S.; visualization, M.C.; writing---original draft, M.C.; writing---review and editing, M.C., T.F., N.-U.B., T.K. and M.S.;}

\funding{This research received no external funding.}

\acknowledgments{N.-U.B. and T.F. acknowledge support from the Polish National Science Center (NCN) under Grant No. UMO-2016/23/B/ST2/00720.}

\conflictsofinterest{The authors declare no conflict of interest.} 

\reftitle{References}

\end{document}